\documentclass[sigconf]{acmart}

\acmSubmissionID{270}
\newcommand{\yuting}[1]{\textcolor{magenta}{}}
\newcommand{\deepak}[1]{\textcolor{yellow}{}}
\newcommand{\jessica}[1]{\textcolor{olive}{}}
\newcommand{\greg}[1]{\textcolor{teal}{}}
\newcommand{\jungdam}[1]{\textcolor{orange}{}}
\newcommand{\yunbo}[1]{\textcolor{blue}{}}
\newcommand{\TODO}[1]{\textcolor{red}{}}
\newcommand{\compIM}[1]{#1}
\newcommand{\limVar}[1]{#1}
\newcommand{\limGen}[1]{#1}

\usepackage{ dsfont }
\usepackage{subcaption}
\usepackage{soul}
\acmJournal{TOG}

\copyrightyear{2023}
\acmYear{2023}
\setcopyright{rightsretained}
\acmConference[SIGGRAPH '23 Conference Proceedings]{Special Interest Group on Computer Graphics and Interactive Techniques Conference Conference Proceedings}{August 6--10, 2023}{Los Angeles, CA, USA}
\acmBooktitle{Special Interest Group on Computer Graphics and Interactive Techniques Conference Conference Proceedings (SIGGRAPH '23 Conference Proceedings), August 6--10, 2023, Los Angeles, CA, USA}
\acmDOI{10.1145/3588432.3591491}
\acmISBN{979-8-4007-0159-7/23/08}

\AtBeginDocument{%
  \providecommand\BibTeX{{%
    \normalfont B\kern-0.5em{\scshape i\kern-0.25em b}\kern-0.8em\TeX}}}

\begin{document}

\title{Simulation and Retargeting of Complex Multi-Character Interactions}

\author{Yunbo Zhang}
\authornote{This work is conducted while authors were working at Meta}
\email{ybzhang3027@gatech.edu}
\orcid{0009-0009-3612-1203}

\affiliation{%
  \institution{Georgia Institute of Technology}
  \country{USA}
}

\author{Deepak Gopinath}
\authornotemark[1]
\email{d_gopinath@apple.com}
\orcid{0000-0002-1023-1297}
\affiliation{%
  \institution{Apple}
  \country{USA}
 }

\author{Yuting Ye}
\email{yuting.ye@meta.com}
\orcid{0000-0003-2643-7457}
\affiliation{%
  \institution{Reality Labs Research, Meta}
  \country{USA}
 }

\author{Jessica Hodgins}
\email{jkh@cmu.edu}
\orcid{0000-0002-1778-883X}
\affiliation{%
  \institution{Carnegie Mellon University}
  \country{USA}
}

\author{Greg Turk}
\email{turk@cc.gatech.edu}
\orcid{0000-0002-3419-6369}
\affiliation{%
  \institution{Georgia Institute of Technology}
  \country{USA}
}

\author{Jungdam Won}
\authornote{corresponding author}
\orcid{0000-0001-5510-6425}
\email{jungdam@imo.snu.ac.kr}
\affiliation{%
 \institution{Seoul National University}
 \country{South Korea}
}


\begin{abstract}
 We present a method for reproducing complex multi-character interactions for physically simulated humanoid characters using deep reinforcement learning. Our method learns control policies for characters that imitate not only individual motions, but also the interactions between characters, while maintaining balance and matching the complexity of reference data. Our approach uses a novel reward formulation based on an \textbf{interaction graph} that measures distances between pairs of interaction landmarks.  This reward encourages control policies to efficiently imitate the character's motion while preserving the spatial relationships of the interactions in the reference motion. We evaluate our method on a variety of activities, from simple interactions such as a high-five greeting to more complex interactions such as gymnastic exercises, Salsa dancing, and box carrying and throwing. This approach can be used to ``clean-up'' existing motion capture data to produce physically plausible interactions or to retarget motion to new characters with different sizes, kinematics or morphologies while maintaining the interactions in the original data.

\end{abstract}

\ccsdesc[500]{Computing methodologies~Physical simulation}
\ccsdesc[500]{Computing methodologies~Motion capture}
\ccsdesc[300]{Computing methodologies~Reinforcement learning}
\ccsdesc[300]{Computing methodologies~Learning from demonstrations}

\keywords{Character Animation, Interactions, Physics Simulation, Physics-based Characters, Reinforcement Learning}

\begin{teaserfigure}
\centering
  \includegraphics[width=0.93\textwidth]{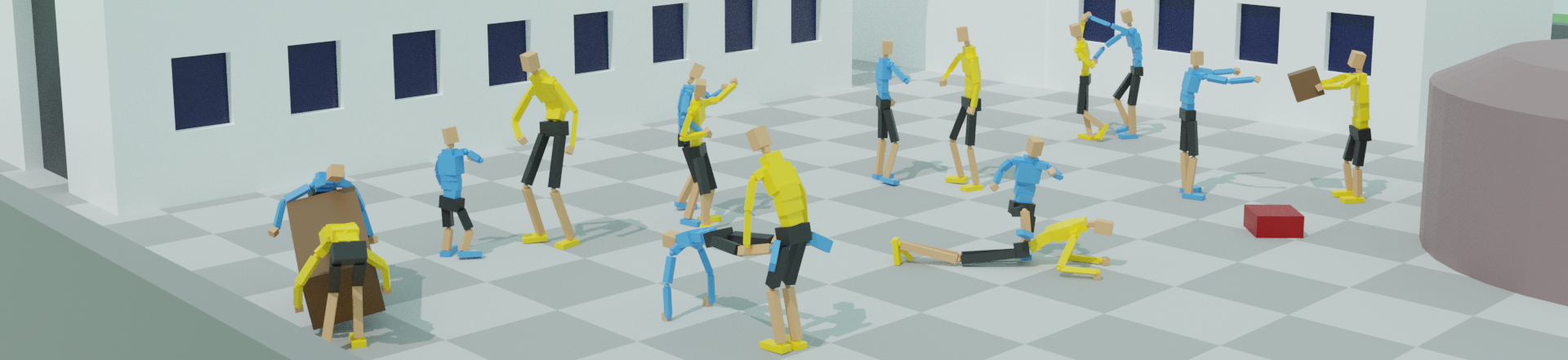}
  \caption{A variety of two-character interactions animated using on our method.}
  \label{fig:teaser}
\end{teaserfigure}

\maketitle

\section{Introduction}



\jessica{do we have a plan for a teaser figure?} \jungdam{Should be done after finishing Blender rendering.}

Physical interactions between people are important elements in daily life.  For example, we can imagine a simple scenario where people greet each other by shaking hands, and a more complex scenario where two people perform physical activities together such as Yoga or Salsa dancing. 
A virtual character capable of reproducing such interactions with either autonomous or user-controlled characters in a believable manner would provide an immersive experience in applications such as computer games, movies, or AR/VR platforms.  

In this paper, we are interested in transferring complex multi-character interactions from reference motions to physically simulated characters. Those characters need to be carefully coordinated in both spatial and temporal domains.  Interactions among physically simulated characters have been studied far less than single characters, in part because it is very challenging to learn controllers for multiple characters interacting with each other.  As with a single character, balance must be maintained, but the interaction constraints also have to be solved simultaneously.  Although some breakthrough results were demonstrated in recent studies~\cite{Haworth:2020, Won:2021:FairPlay, Siqi:2022:DeepMindFootBall}, the complexity of the demonstrated interactions are still far from what people routinely perform in daily life. 



We demonstrate a novel learning-based method that provides a physics-based retargeting of complex interactions for multiple characters.  More specifically, given reference motions that capture interactions between people, we learn control policies (a.k.a. controllers) of simulated characters via deep reinforcement learning that imitate not only the motion of the individuals but also the interactions between them.  Our learned policies can produce plausible and semantically equivalent interactions when the sizes and kinematics of the characters are varied significantly.   If the size of the simulated characters match those in the original motion capture data, the resulting motion is almost indistinguishable from the reference data and any errors from the capture process are eliminated by ensuring that the interactions are now physically plausible.  To solve the challenges in learning multi-character interactions, we develop new rewards based on an \textbf{interaction graph} (IG) which measures distances between pairs of specified locations on the characters, and in particular reflects between-character distances. Rewards based on the IG enable control policies to efficiently deploy complex interactions for physically simulated characters while preserving the semantics of the interactions (i.e. spatial relationship) included in the reference data.  In our formulation, manual annotation of the interaction in each motion is not necessary except for choosing a set of general interaction landmarks that work for a variety of scenarios.
\jessica{it seems like we should lead with the ability to modify the motion, not that we don't break the original mocap.   In the last sentence, is physically plausible the right phrasing?  Rather than ``natural'' or just ``plausible''.  It is going to be physically correct, since it is simulated?} \jungdam{My point was that our method does not break the original semantics rather than motions. I think your point also makes sense, so I added an extra sentence at the end. I also edited "physically plausible".} \jessica{it still seems to me that it would be more impressive to lead with our ability to change the characters.   Just maintaining the contacts that are present in the original mocap (presumably) seems less impressive.  If we make this change, we should make it in the abstract as well.} \yuting{We could also emphasize there is no need to manually annotate the interaction in each motion, except for choosing a set of general interaction landmarks that work for a variety of scenarios. Not sure if we want to mention self-interaction when changing size.} \jungdam{To reflect the comments, I reorganized the intro by mixing and editing the previous and current paragraphs.}

\jessica{ Here is some proposed text for the last half of the paragraph above}
\jessica{Our method provides a physics-based retargeting of complex interactions for multiple characters. Our learned policies can produce plausible and semantically equivalent interactions when the sizes and kinematics of the characters are varied significantly.   If the size of the simulated characters match those in the original motion capture data, the resulting motion is almost indistinguishable from the reference data and the interactions are now physically plausible. } \jungdam{Revised mentioned above.}

To show the effectiveness of our method, we record motions that include multi-person interactions at varying levels of difficulty, and test our method with motions that are composed of simple interactions such as a high-five or other greetings,  as well as complex interactions such as gymnastic exercises, Salsa dancing, and box moving/throwing. We demonstrate the generality of our system by reproducing interactions not only for simulated characters with different body dimensions than the motion capture subjects,  but also for a robot with a different kinematic structure.  Finally, we run comparison and ablation studies that justify each choice in the system design.  

\section{Related work}

We first review papers synthesizing multi-character interaction for kinematic approaches, which inspired our dynamic formulation.  We then review recent progress in control of physically simulated characters via deep reinforcement learning.  

\subsection{Multi-character Interactions for Kinematic Characters}


Most approaches for creating or editing multi-character interactions among kinematic characters  are data-driven methods, which means that appropriate motion capture data should be obtained in advance.  A popular line of work is based on optimization, where the basic idea is to optimize individual character motions with spatio-temporal constraints~\cite{Liu:2006, Kwon:2008}, game theory~\cite{Shum:2007, Shum:2008b, Shum:2012, Wampler:2010} so that the optimized motions have newly synthesized interactions.  These methods are suitable for synthesizing motions having sparse interactions, however, the optimization quickly becomes intractable as the complexity of the interactions increases, so it is not suitable for synthesizing dense and complex interactions.  Another approach is patch-based methods, where a patch includes a short interaction of multiple characters~\cite{Lee:2006:MotionPatch, Shum:2008:InteractionPatch, Yersin:2009:CrowdPatch, Hyun:2013:Tile, Won:2014:Previs}.  For this work, motion capture data where multiple actors are recorded simultaneously is required. New motions can be synthesized by connecting boundaries of multiple patches, thus creating multiple interactions that were not performed together in the original data.  

Methods for adapting existing interactions to new environments and characters have also been studied~\cite{Kim:2009:SyncEdit,Ho:2010:Spatial,Al-Asqhar:2013:RelDescriptor,Kim:2014:Crowd,Jin:2018:AuraMesh,ho:2014:multi-res,Kim-2021-Interactive}.  The key idea is to define an interaction descriptor that encodes the spatial and temporal relationship, then to edit the motions while minimizing the semantic difference between the original motion and the edited motions where the difference is measured by the descriptor.  This idea has also been used to synthesize hands interacting with objects~\cite{Zhang:2021:ManipNet}.  Our state representation and reward function for deep reinforcement learning are inspired by one of these descriptor-based approaches~\cite{Ho:2010:Spatial}, where they construct an interaction graph by connecting edges among pre-specified markers on the body surface. By utilizing deep reinforcement learning and a novel formulation to measure interaction graph similarities, our method can be applied to dynamic characters having different body shapes instead of generating kinematic interaction motions as was done in~\cite{Ho:2010:Spatial}. 

\subsection{Physically Simulated Characters and Interactions}

In many cases, what we refer to as \textit{interaction} between different characters means physical interaction where physical forces occur between the characters at contacts.  By incorporating physics simulation into the motion of the characters, those physical interactions can be synthesized in a plausible manner.  Multi-character interactions have been created by solving a quadratic programming problem where the equations of motion for the entire dynamical system are used as either hard or soft constraints~\cite{Mordatch:2012:CIO,Otani:2017,Vaillant:2017}.  Although cooperative multi-character interactions could be synthesized by these methods without using reference data, the generated motions are typically slow and less-dynamic due to the quasi-static assumption in their optimization formulation, and they require frame-level specification of all interactions in advance. Combining deep reinforcement learning (DRL) and motion capture data has allowed several breakthroughs in learning imitation controllers~\cite{Peng:2018:DeepMimic,Chentanez:2018,Park:2019, Bergamin:2019:DReCon,Won:2020:ScaDiver,Levi:2021:SuperTrack,Peng:2021:AMP}, learning reusable motor skills~\cite{Peng:2019:MCP,Merel:2019:b,Peng:2022:ASE, Won:2022:cVAE,yao2022controlvae}, and motion tracking~\cite{ye2022neural3points,winkler2022questsim}.  Although there also have been some studies synthesizing dynamic interactions with objects or other characters~\cite{Merel:2020:CatchCarry,Haworth:2020,Won:2021:FairPlay, Siqi:2022:DeepMindFootBall}, the complexity of the demonstrated interactions are still not comparable to what people routinely perform in daily life. \compIM{In addition, each of these works developed a task-specific reward function to enforce interactions between multiple entities.} In this paper, we aim to synthesize \compIM{various types of }spatially and temporally dense interactions for full-body humanoid characters that are physically simulated.  This problem is especially challenging because the motor skills must be sophisticated enough to perform those complex interactions while remaining robust enough to maintain balance.

\section{Method}

Our goal is to build controllers that enable physically simulated characters to perform complex physical interactions with each other. For each behavior, we take a reference motion capture clip representing the desired multi-character interaction and produce controllers that enable the simulated characters to mimic those interactions. 
\jessica{by saying "a set of" above it seems to imply that more than one motion is used at a time.   That's not true, correct? I'm also not sure that calling it an example based approach is right as again it sounds like the examples are merged in some way.   I would say ``For each behavior, we take a reference motion capture clip representing the desired behavior ...''}\yunbo{Fixed} Our goal is to generate character interactions that are \textit{semantically} similar to those present in the reference motions.  To achieve this, we use multi-agent deep reinforcement learning where the states and rewards are designed based on spatial descriptors inspired by~\cite{Ho:2010:Spatial}. Different from ~\cite{Ho:2010:Spatial} where only kinematic motions are generated, our method can be applied to dynamic characters having dramatically different body shapes from the captured actors. 

\subsection{Environment}
Our characters are modeled as articulated rigid body objects by following~\cite{Won:2020:ScaDiver}.  Each character has 22 links and 22 joints, where each joint has three degree-of-freedom and is actuated by stable-PD servos~\cite{2011-Tan-SPD} given target joint angles.  We used an open-source framework~\cite{Won:2020:ScaDiver} to implement and simulate our characters. 

\subsection{Problem Formulation}
We formulate the problem as a multi-agent Markov Decision Process (MDP). Consider $k$ controllable agents, we define the tuple $\{S, O_1\cdots O_k,A_1\cdots A_k,R_1\cdots R_k,T,\rho\}$ where $S$ is the entire state of our environment, $O_i$ and $A_i$ are the observation and action of $i$-th agent, respectively. The reward function $R_i : O_i\times A_i \rightarrow \mathbb{R}$ evaluates the quality of the current state and action of $i$-th agent,  the environment is updated by the transition function $T : S\times A_1\times \cdots \times A_k \rightarrow S$ given a set of actions performed by all the agents, and $\rho : S \rightarrow [0,1]$ is the probability distribution of the initial states.  We aim to learn a set of optimal control policies $\{\pi_i | i=1 \cdots k\}$ that maximizes average expected return $\mathbb{E}\left[\sum_{t=0}^{T}\gamma^tr_{i,t}\right]$ for each agent, where $\gamma \in (0,1)$ is the discount factor that prevents the sum from being infinity. 

\subsection{Interaction Graph}
\begin{figure}[H]
    \centering
    \includegraphics[width=0.8\columnwidth]{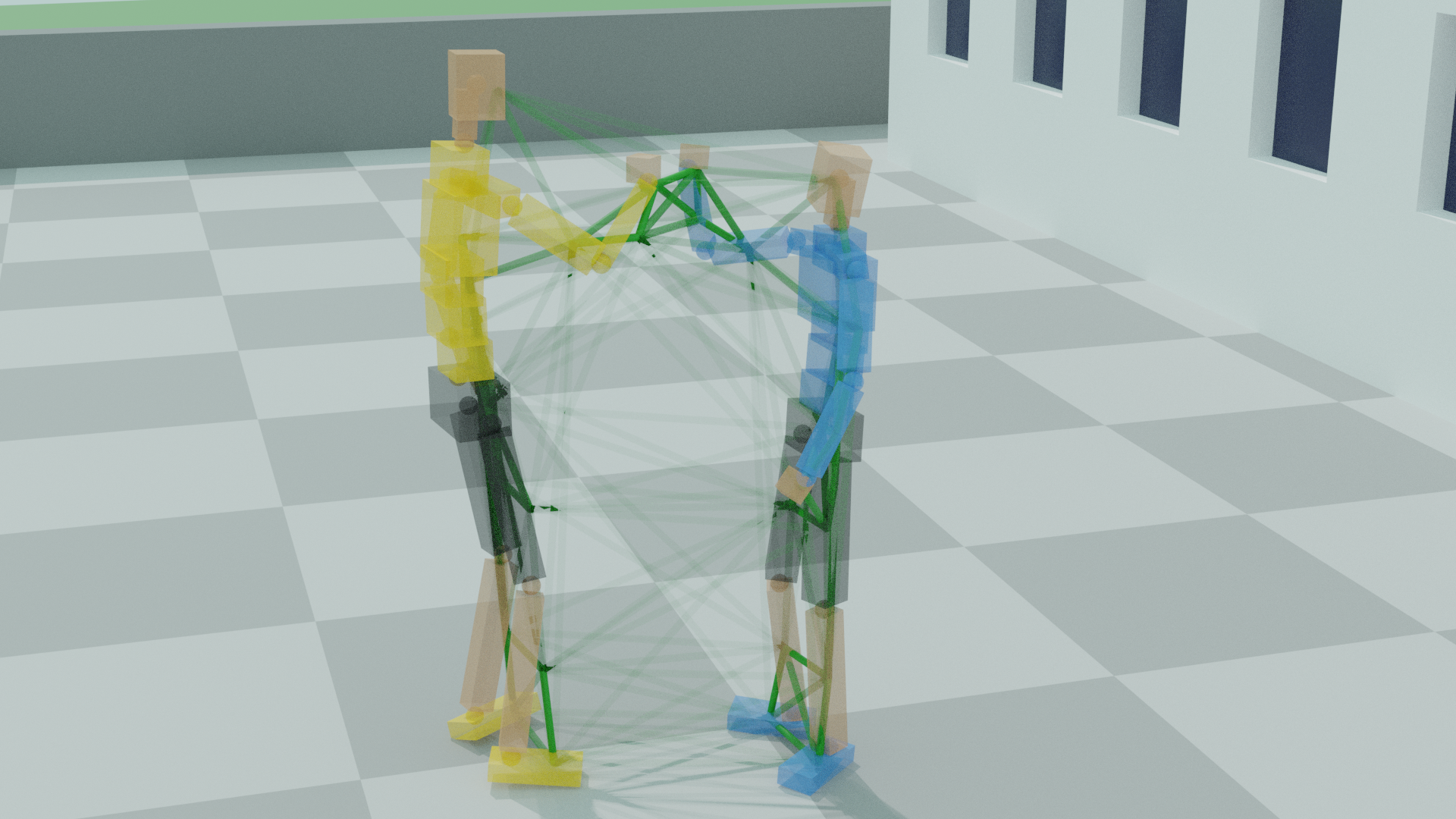}
    \caption{Interaction Graph of the reference characters. Higher opacity on an edge indicates a higher weight for the edge when computing the reward.}
    \label{fig:IG_two_characters}
\end{figure}

To better describe the semantics of the interaction happening between agents (or between an agent and an object) during the motion, we define the notion of an Interaction Graph (IG), a graph-based spatial descriptor where the information on interactions is stored in its vertices and edges. This idea is inspired by~\cite{Ho:2010:Spatial}. To construct an interaction graph, we first place a collection of markers on salient locations on each character (see Figure~\ref{fig:IG_two_characters}).   Fifteen  markers are placed in total for each character, where three markers are on each limb in the vicinity of joint locations, one on the pelvis, one on the torso, and one on the head.  These markers will be considered as the nodes of the graph, each of which is associated with a $6$-dimensional vector $n_i = (p_i,v_i)\in\mathbb{R}^6$, where $p_i\in\mathbb{R}^3$ is the position of the vertex and $v_i\in\mathbb{R}^3$ is the velocity of the vertex.  For example, a total of 30 vertices will be used for interactions associated with two characters (see Figure~\ref{fig:IG_two_characters}). On every time step, we perform a Delauney Tetrahedralization over all the vertices based on the spatial distances between pairs of markers to get a compact collection of edges connecting the vertices. Each edge is assigned a feature vector $e_{ij} = (p_{ij}, v_{ij})\in\mathbb{R}^6$ that encodes the relative relationship between the two vertices, where $p_{ij}=p_j-p_i\in\mathbb{R}^3$ and $v_{ij}=v_j-v_i\in\mathbb{R}^3$ are the positional and velocity components of the edge features.  


The example interaction graph in Figure \ref{fig:IG_two_characters}  includes both edges connecting nodes on a single character and edges connecting nodes on different characters. The edges within the character help maintain the motion quality of an individual character, while the edges between the characters act as guides for maintaining the relative position of the body parts of the two characters. Details are discussed later in section \ref{sec:reward_design}.

There is a major difference between how we compare two spatial descriptors in the interaction graph and how they are compared in the Interaction Mesh (IM) in~\cite{Ho:2010:Spatial}. 
We perform edge-level (i.e. distance) computation whereas IM computes volumetric deformation on a tetrahedron. We further augment the state of an edge with velocities as they are crucial for a physics simulation. Given the input reference motions clips, we build and store such an IG to capture the spatial relationship across the agents and object at each time-step.  

\subsection{Reward Design}
\label{sec:reward_design}

We choose to measure the interaction similarity in two ways:  an \textit{edge-weighting function} that highlights the importance of interaction regions in the graph and an \textit{edge-similarity function} that measures the similarity between two IGs with the same connectivity.

 For the following similarity measurement, we make use of two interaction graphs $G^{sim}$ and $G^{ref}$ with the same connectivity, one from the simulated environment, the other from the reference motion clips. The connectivity of both graphs is the same as computed on the reference motions using the above mentioned method. The interaction graph we defined is a set of spatial descriptors that encode the relative formation among the vertices in the graph. 

\subsubsection{Edge Weighting Function}
We are guided by the intuition that instances where two body parts are close or in contact with each other are particularly important for multi-character interactions.  We define a function that dynamically assigns different weights for each edge according to its relative importance to the others.  More specifically, for an edge connecting vertices $i$ and $j$, the weight for the edge $w_{ij}$ is defined as:

\begin{equation}
    w_{ij} = 0.5*\frac{\exp{\left(-k_{w}\|p^{sim}_{ij}\|\right)}}{\sum_{ij}\exp{\left(-k_{w}\|p^{sim}_{ij}\|\right)}}+0.5*\frac{\exp{\left(-k_{w}\|p^{ref}_{ij}\|\right)}}{\sum_{ij}\exp{\left(-k_{w}\|p^{ref}_{ij}\|\right)}}
    \label{eq:edge_weighting_function},
\end{equation}
where $k_{w}$ controls how sensitive the weighting function is with respect to the distance of the edges. The first term gives more attention to an edge if the two nodes in the simulation are close to each other, the second term makes sure an edge in the reference motion gets more attention when its nodes stay close. In practice, we found the second term alone is enough for most of our experiments, and we only use the first term for a few examples where it improves the performance.

Normalizing the weights allows our reward function to adapt to various interaction scenarios.  For example, when two characters are far away from each other, the edges connecting those two characters do not contribute much to the reward while the edges connecting vertices within individual characters become important.  
On the other hand, when the two characters are close to each other, some of the connections between their body parts will have large weights.  This adjustment based on proximity allows the body parts that are not associated with the close interactions to remain close to the original motion. \jessica{my fixes to the above paragraph may have broken it -- Yunbo should reread.} \yuting{I would like to see something like this in the intro :)}

\subsubsection{Edge Similarity Function}
Given two interaction graphs $G$ and $G'$, we design a distance function measuring their differences. Our distance function measures position and velocity similarity of the two different formations by comparing all corresponding edges in the two graphs.



\subsubsection{Positional Graph Similarity}
To compare the positional graph similarity between two graphs, we separately consider the similarity of the two graph edges connecting each individual character $E_{self}$ (self-connections) and between characters $E_{cross}$ (cross-connections).  The discrepancy of each edge is computed as follows:
\begin{equation}
    err_{self, ij} = \|\frac{p^{sim}_{ij}-p^{sim}_{T,ij}}{\|p^{sim}_{T,ij}\|} - \frac{p^{ref}_{ij}-p^{ref}_{T,ij}}{\|p^{ref}_{T,ij}\|}\|
\end{equation}
where $p^{sim}_{T,ij}$ and $p^{ref}_{T,ij}$ are edges computed from the first frame of the motion sequence  for both simulation and reference motions. In all the reference motion sequences, the motion capture actors are instructed to start in a T-pose. In other words, we first compute the deviation of an edge from its corresponding T-pose edge, it is then normalized by the length of the T-pose edge.  Finally, we compute the difference between the two deviations, one for the simulated characters and the other for the reference motion clips.  Note that this similarity measurement is not sensitive to specific body sizes and proportions due to the normalization of the deviations.
This formulation is also similar to measuring the Laplacian coordinate difference of all graph nodes between simulation and reference in that they both try to maintain the similarity of the local structure between two graphs, but our formulation gives direct measure to the edge similarity that captures the interaction. 

\TODO{Write some notes on similarity to Laplacian operator}
\yunbo{Added notes on Similarity to Laplacian operator.
}


It is challenging to define a reference edge length for the cross-connections because the variance can be extremely high.  For example, imagine that the two characters are standing 10m apart versus 0.1m.  Instead, we directly penalize the difference in the edge length and direction so that the same cross-connection similarity can be applied to various motions:


\begin{equation}
    err_{cross, ij} = 0.5*\frac{\|p^{sim}_{ij}-p^{ref}_{ij}\|}{\|p^{sim}_{ij}\|} + 0.5*\frac{\|p^{sim}_{ij}-p^{ref}_{ij}\|}{\|p^{ref}_{ij}\|}
    \label{eq:cross_err}
\end{equation}
where we normalize the difference by the lengths in the simulation and the reference clips, respectively, then average them so that the similarity becomes symmetric.  This symmetry also enables the error to be used for characters having different body shapes.  The total error for positional graph similarity is then the sum of the two error terms from all edges:
\begin{equation}
    err_{pos\_graph} = \sum_{ij\in E_{cross}}{w_{ij}err_{cross, ij}} + \sum_{ij\in E_{self}}{w_{ij}err_{self, ij}} 
\end{equation}

\subsubsection{Velocity Graph Similarity}
To measure the velocity discrepancy between graphs, we simply measure the difference of velocity of all edges in simulation and reference as:
\begin{equation}
    err_{vel\_graph} = \sum_{ij\in E_{cross}\cup E_{self}}w_{ij}\|v^{sim}_{ij}-v^{ref}_{ij}\|
\end{equation}
In contrast to the positional similarities, we observed that the velocities of the graph vertices do not vary much when modifying the body size/proportion of the simulated characters. Thus we do not perform any velocity normalization. We also do not separate the velocity similarities by edge type because we did not find any benefit in doing so.

\subsubsection{Final Reward Design}
We define our reward function based on the errors computed from the interaction graphs.  In addition, we add two more error terms measuring the tracking of the root joint and center-of-mass, which are frequently used in learning imitation controllers for physically simulated characters.  As a result, our reward function is composed of four terms
\begin{equation}
    r = r_{pos\_graph}\cdot r_{vel\_graph}\cdot r_{root}\cdot r_{com}
\end{equation}
\begin{equation} 
    \begin{split}
        r_{pos\_graph} = \exp(-k_1*err_{pos\_graph})\\
        r_{vel\_graph} = \exp(-k_2*err_{vel\_graph})\\
        r_{root} = \exp(-k_3*err_{root})\\
        r_{com} = \exp(-k_4*err_{com})
    \end{split}
\end{equation}
where $r_{pos\_graph}$, $r_{vel\_graph}$ measure the difference between the two interaction graphs and $r_{root}$, $r_{com}$ encourage the tracking of the root joint and the center-of-mass projected on the ground, $k_1,\cdots,k_4$ are the sensitivities of the terms, respectively. The errors for the tracking are defined as follows:
\begin{equation}
\begin{split}
    err_{root} = w_p\|\bar{p}_{sim}-\bar{p}_{ref}\|^2+w_q\|log(q_{sim}^{-1}\cdot q_{ref})\|^2 +\\ w_v\|\bar{v}_{sim}-\bar{v}_{ref}\|^2 + w_\omega\|\omega_{sim}-\omega_{ref}\|^2
\end{split}
\end{equation}
\begin{equation}
    err_{com} = w_{com,x}\|x_{sim}-x_{ref}\| + w_{com,\dot{x}}\|\bar{x}_{sim}-\dot{x}_{ref}\|
\end{equation}
 where $\bar{p},\bar{v}$ are the position and velocity of the root joint excluding the height components, and $q$,$\omega$ are the orientation and angular velocity of the root joint, respectively, $x$ and $\dot{x}$ are the center-of-mass position and velocity of the simulated character excluding their height components.  $w_p$, $w_q$, $w_v$, $w_w$, $w_{com,x}$, and $w_{com,\dot{x}}$ are the relative weights of the terms.  Note that we ignore the height components of the linear positions and velocities so that the relevant errors are not directly affected by the absolute size of the character. \compIM{In contrast to \cite{Ho:2010:Spatial}, where tetrahedron volumes are used to measure similarities of meshes, our edges-based reward is more sensitive to point-to-point physical interactions. In addition, it is not trivial to design an adaptive weight function in a volume-based setting, which ensures the motion quality of the individual characters is preserved, even when characters are far apart, making our reward a good substitute for motion imitation.}



\subsection{Observation and Action Spaces}

The observation space of our environment is inspired by the design from prior work \cite{Won:2020:ScaDiver, Won:2021:FairPlay} where the observation of an agent $o_i = (o_{sim},o_{ref})$ consists of the states of the simulated characters and objects, which are computed from the simulation and the reference motion clips.  For the simulated observation space $o_{sim} = (o_{sim,self},o_{sim,other},o_{sim,object})$, we include the position, orientation, linear and angular velocity for each link of the characters and the objects. To make sure the state is invariant to the global position and orientation of the agent, all values are transformed to the facing frame of the controlled character. The facing frame of the character is computed by projecting the global transformation of the character root to the ground.

The reference observation $o_{
ref}=(o^{0}_{ref},o^{0.05}_{ref},o^{0.15}_{ref})$ contains the reference information $0$, $0.05$, and $0.15$ seconds in the future. For each future reference observation frame $o^{*}_{ref}=(o^{*}_{ref,self},o^{*}_{ref,other}\\,o^{*}_{ref,object})$, we include the position, orientation, linear and angular velocity for each link of the characters and the objects in the facing frame of the reference character.  \jungdam{TODO: This should be written more precisely.}\yunbo{Added more detail}

Our action $a$ is the change of pose $\Delta q$ from the pose $q_{ref}$ given the reference frame at each time-step.  A new reference pose $q_{ref}+\Delta q$ (i.e. a set of joint angles) is given to the stable PD servos attached to our simulated character and then joint torques are computed accordingly.
\section{Results}
In this section, we show that our formulation can be applied to a variety of motions with multiple characters and objects. By dynamically adjusting the weights, our method focus on the adjustments to the motion on the physical interactions.  This approach results in higher quality motion than existing work in scenarios with complex interactions. Further, our formulation is able to preserve interaction when the body size, kinematics, and skeleton of the characters differ from the reference motion sequences.

\subsection{Experiment Setup}
The structure of our policy follows a encoder-decoder style as presented in~\cite{Won:2021:FairPlay}, where the encoder is a fully connected neural network with two hidden layers with $256$ and $128$ units respectively. The encoder takes the full observation and projects it onto a $32$ dimensional latent vector $z$. The decoder is another fully connected network with two hidden layers with $256$ units, and it takes as input the concatenated vector $z_{decoder} = (o_{sim,self},z)$ and outputs the action of the policy. To speed up the learning for all of the experiments below, we pre-train an imitation policy of a single character on sequences that can be performed without a partner (e.g. high five, greetings, and push ups). When training an interaction-graph based policy, we reuse the pre-trained decoder and allow its weights to be updated during the training. The decoder is reusable because the latent dimensions are unchanged.  The encoder trained simultaneously with the pre-trained decoder is not reusable due to differences in input dimensions.  This design makes it easier for the policy to maintain balance at the initial phase of learning, and therefore results in faster training. The training time of a policy varies based on the difficulty of the sequence. For easier sequences, it takes about 300 million to 500 million samples to train one policy. For harder sequences, it could take more than 2 billion samples to train a policy. All experiments are run using $640$ CPUs and take from $3$ days to $9$ days to train a policy based on the sequence difficulty. \jungdam{XXX should be resolved.}\yunbo{Resolved by specifying the sample sizes, compute resources, and training time.} 

\subsection{Human-Human Interaction}
In the human-human interaction scenarios, we aim to show that our method is capable of reproducing imitation policies with similar motion quality as existing works such as \cite{Peng:2018:DeepMimic, Won:2020:ScaDiver, Levi:2021:SuperTrack} while the interaction is better preserved. We show a variety of scenarios ranging from sparsely interacting motions to continuously interacting motions between the two human characters. 
\subsubsection{Light Interaction}
Figure~\ref{fig:light_interaction_greeting} and ~\ref{fig:light_interaction_jumpover} shows light physical interactions. In \textit{Rapper-Style Greetings}, the two characters  touch their hands, elbows, shoulders, and legs in sequence to greet each other, an action which has been shown in many hip-hop music videos~\cite{video:hip-hop-greetings}. In \textit{Jumpover}, one character jumps over the other character. In these scenarios, physical interactions are of short duration with little physical forces, and the interactions are well-preserved semantically when the interacting body parts are close enough with the right timing. \jessica{can't parse previous sentence}\yunbo{Resolved.} 

\subsubsection{Heavy Interaction}

Figure~\ref{fig:heavy_interaction_lift_pushup}, ~\ref{fig:heavy_interaction_salsa_grasp}, and ~\ref{fig:heavy_interaction_salsa_support} shows physical interactions where significant forces occur between the two characters. The \textit{Lift-Pushup} example includes interactions where one character needs to lift the other character's legs while that character is performing a push-up exercise. In the first salsa dancing motion (\textit{Salsa Grasping}), two character's hands are grasped together to form a loop for one character to go under. In another salsa dancing motion (\textit{Salsa Support}), one character needs to support the other while they lean backward. This type of interaction is more challenging than the light interactions because the two simulated characters need to perform highly coordinated motions with force exchange.  For example, the character performing a push-up would not be able to imitate the reference motions successfully unless his legs are grasped by the other character. Furthermore, these heavy interactions make maintaining balance difficult because significant forces are applied between the two characters. Our method was able to imitate these challenging motions successfully as shown in Figure~\ref{fig:heavy_interaction_lift_pushup}. Because our characters do not have fingers, we mimic the grasp by adding weld constraints in the physics simulation. More specifically, we examine the reference motion and label a sequence of grasping windows to identify when grasping should be presented on specified body pairs at certain moment. During simulation, when a character's hand is close to a body part it should grasp at that moment, weld constraints between the two body parts are made temporarily. The constraint is removed when the grasping window is over, representing the character releasing their hands. Other than hand grasping, we did not use any weld constraints, all the complex physical interactions emerged during the learning process.


 These results show that our formulation allows for the control policies to be aware of the relative formation among various body parts regardless of the interaction duration, and to preserve those formations in the simulation. 

\subsection{Human-Object Interaction}

\begin{figure}[H]
    \centering
    \begin{subfigure}{0.45\columnwidth}
        \centering
        \includegraphics[height=3.5cm]{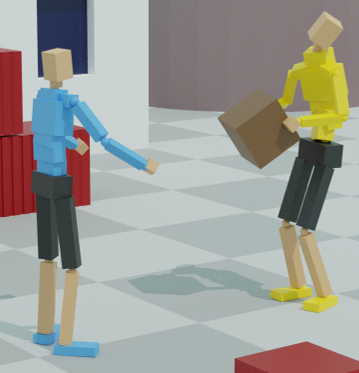}
        \caption{Box Throw and Catch}
        \label{fig:human_object_interaction_box_throw}
    \end{subfigure}
    \begin{subfigure}{0.45\columnwidth}
        \centering
        \includegraphics[height=3.5cm]{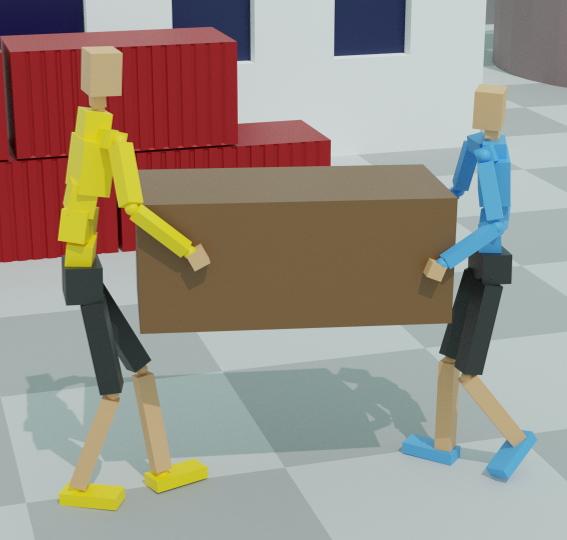}
        \caption{Box Carry}
        \label{fig:human_object_interaction_box_carry}
    \end{subfigure}
     \begin{subfigure}{0.45\columnwidth}
        \centering
        \includegraphics[height=3.5cm]{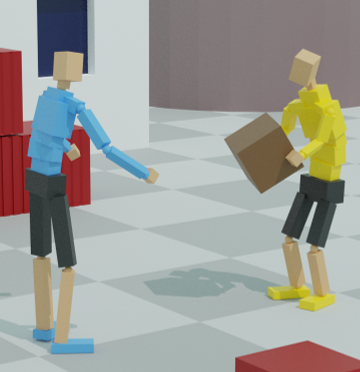}
        \caption{Scaled Box Throw and Catch}
        \label{fig:scaled_human_object_interaction_throw}
     \end{subfigure}
     \begin{subfigure}{0.45\columnwidth}
        \centering
        \includegraphics[height=3.5cm]{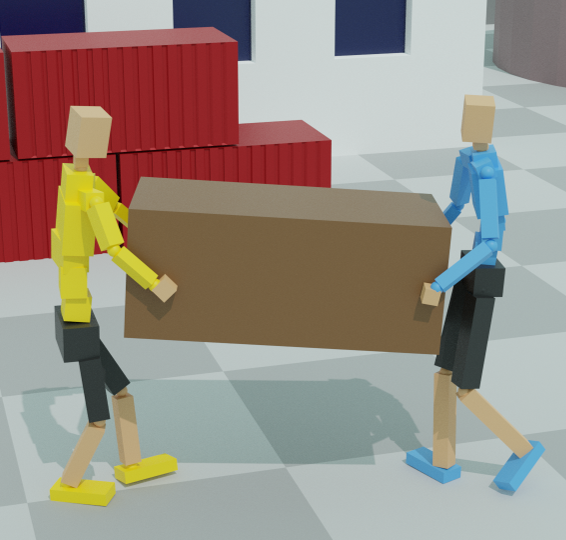}
        \caption{Scaled Box Carry}
        \label{fig:fig:scaled_human_object_interaction_carry}
     \end{subfigure}
    \caption{Human-Object Interaction}
    \label{fig:human_object_interaction}
\end{figure}

\begin{figure}[H]
    \centering
    \includegraphics[width=\columnwidth]{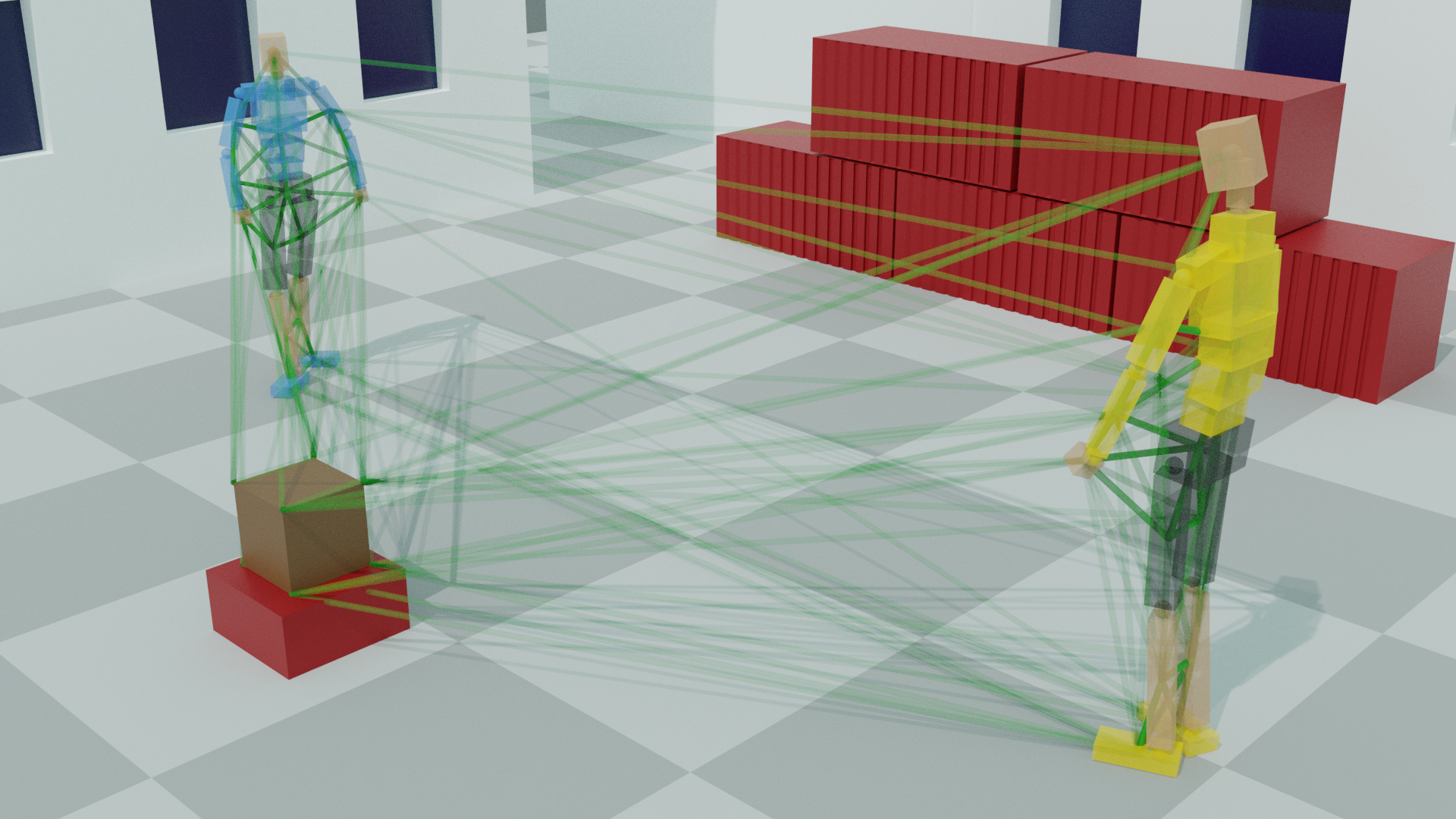}
    \caption{An Interaction Graph that includes an object (the top box) as well as the characters.}
    \label{fig:box_throwing_interaction_graph}
\end{figure}

We further demonstrate that our formulation can also handle human-object interactions where the objects are passively simulated. Figure ~\ref{fig:human_object_interaction_box_throw} and ~\ref{fig:human_object_interaction_box_carry} show the two motions: One includes interactions where two persons are throwing and catching a small box repeatedly, the other includes interactions where two persons are lifting and moving a large box. For both motion sequences, we place an extra marker on every vertex of the box (i.e. 8 markers in total) when constructing the interaction graph. For the edges connecting the characters to the box, we use the reward formulation in Equation ~\ref{eq:cross_err} to measure the discrepancy between simulation and reference. In addition, we choose to remove all edges connecting the markers on the box because their relative distances will stay constant throughout the motion. The resulting graph is shown in Figure \ref{fig:box_throwing_interaction_graph}. The control policies learned with those additional markers successfully reproduce both hand-object interaction scenarios, which shows the generality of our method.

\subsection{Retargeting to different body sizes}
Our graph-based formulation is robust to the change in body conditions because we compute features in a normalized manner. As a result, the interactions in the same reference motions can be applied to simulated characters that have completely different body dimensions from the actors in the reference motions. 

\begin{figure}[H]
\centering
    \begin{subfigure}{0.45\columnwidth}
        \centering
        \includegraphics[height=3.5cm]{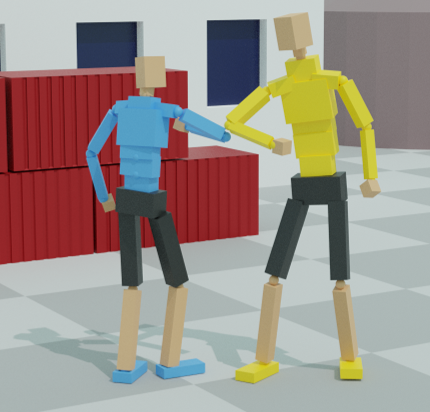}
        \caption{Rapper-Style Greeting}
        \label{fig:light_interaction_greeting}
    \end{subfigure}
    \begin{subfigure}{0.45\columnwidth}
        \centering
        \includegraphics[height=3.5cm]{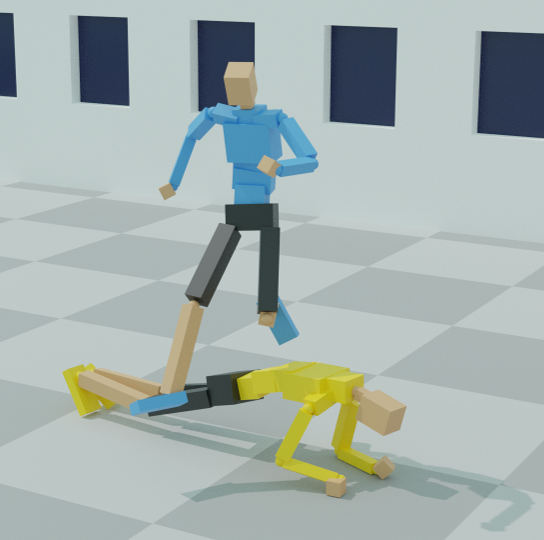}
        \caption{Jumpover}
        \label{fig:light_interaction_jumpover}
    \end{subfigure}
     \begin{subfigure}[b]{0.45\columnwidth}
        \centering
        \includegraphics[height=3.5cm]{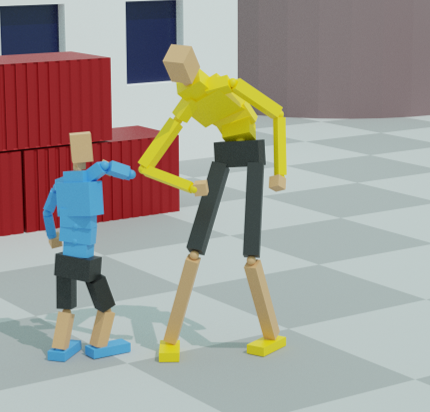}
        \caption{Scaled Rapper-Style Greeting}
        \label{fig:scaled_light_interaction_greetings}
     \end{subfigure}
     \begin{subfigure}[b]{0.45\columnwidth}
        \centering
        \includegraphics[height=3.5cm]{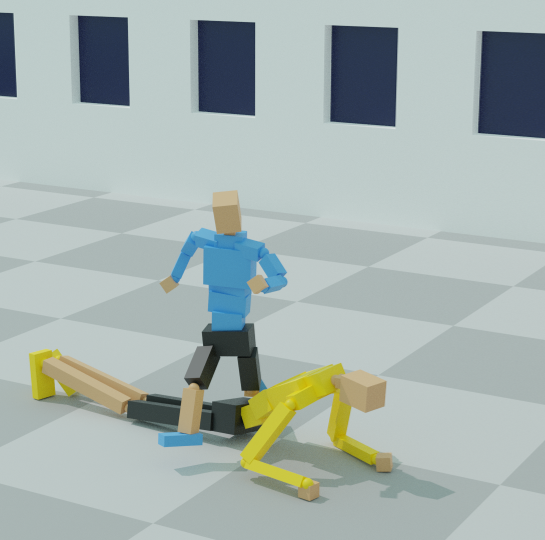}
        \caption{Scaled Jumpover}
        \label{fig:scaled_light_interaction_jumpover}
     \end{subfigure}
    \caption{Light Interaction}
    \label{fig:light_interaction}
\end{figure}

Figure \ref{fig:scaled_light_interaction_greetings}, \ref{fig:scaled_light_interaction_jumpover} demonstrates motions that include light interaction. In both   sequences, we scale all limbs of the yellow and blue characters by a factor of $1.3$ and $0.5$, respectively, so the yellow character is almost 2 times taller than the blue character. The scaled characters are trained using our framework to track the reference motion. In the \textit{Rapper-style Greeting} motion, for example, we see the taller character deliberately bends down to reach their hand, elbow, and shoulder to the shorter character when the interaction happens, and they straighten back after they finish the interaction. Similarly, the taller character lowers their waist when the shorter character jumps over their back in the \textit{Jumpover} motion.

\begin{figure*}[!t]
     \begin{subfigure}{0.3\textwidth}
         \centering
            \includegraphics[height=4cm]{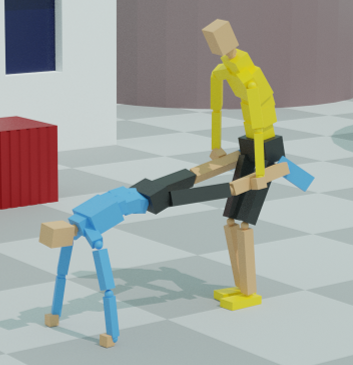}
            \caption{Lift Pushup}
         \label{fig:heavy_interaction_lift_pushup}
     \end{subfigure}
     \begin{subfigure}{0.3\textwidth}
         \centering
            \includegraphics[height=4cm]{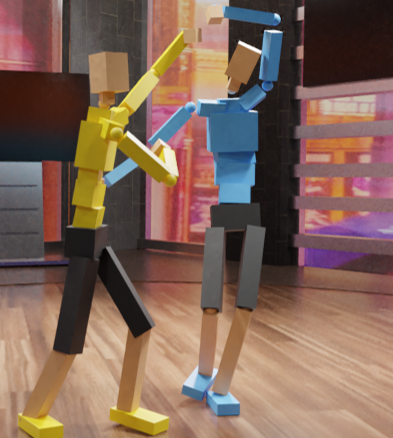}
            \caption{Salsa Grasping}
         \label{fig:heavy_interaction_salsa_grasp}
     \end{subfigure}
     \begin{subfigure}{0.3\textwidth}
         \centering
            \includegraphics[height=4cm]{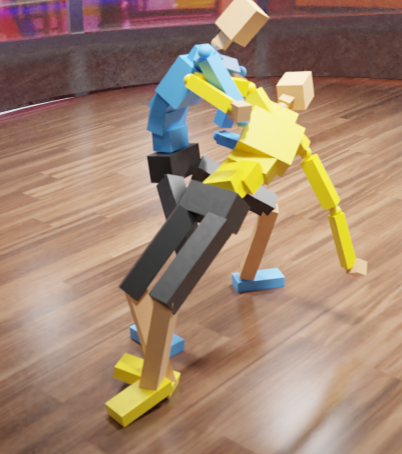}
            \caption{Salsa Support}
         \label{fig:heavy_interaction_salsa_support}
     \end{subfigure}
     \begin{subfigure}{0.3\textwidth}
        \centering
        \includegraphics[height=4cm]{figs/different_scale/IG_scaled_lift_pushup.png}
        \caption{Scaled Lift Pushup}
        \label{fig:scaled_heavy_interaction_lift_pushup}
     \end{subfigure}
     \begin{subfigure}{0.3\textwidth}
        \centering
        \includegraphics[height=4cm]{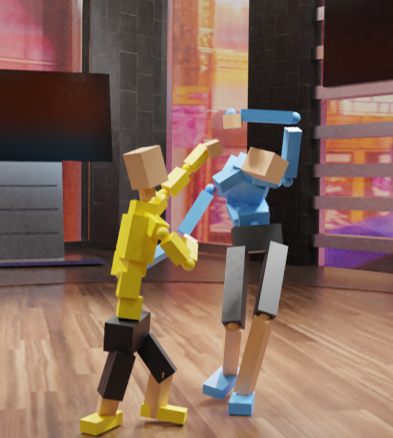}
        \caption{Scaled Salsa Grasping}
        \label{fig:scaled_heavy_interaction_salsa_grasp}
     \end{subfigure}
     \begin{subfigure}{0.3\textwidth}
        \centering
        \includegraphics[height=4cm]{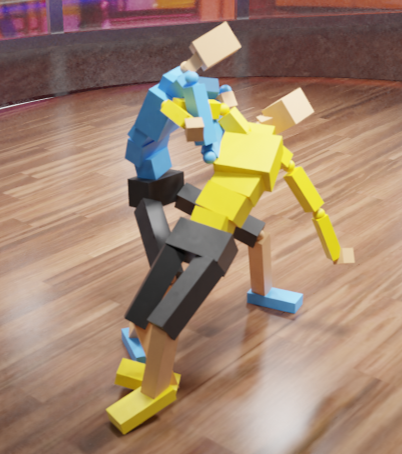}
        \caption{Scaled Salsa Support}
        \label{fig:scaled_heavy_interaction_salsa_support}
     \end{subfigure}
    \caption{Heavy Interaction}
    \label{fig:heavy_interaction}
\end{figure*}

Learning how to transfer forces via physical interactions is crucial to imitating motions including heavy interaction as in Figure~\ref{fig:scaled_heavy_interaction_lift_pushup}, \ref{fig:scaled_heavy_interaction_salsa_grasp},and \ref{fig:scaled_heavy_interaction_salsa_support}. For the \textit{Lift-Pushup} motion (Figure ~\ref{fig:scaled_heavy_interaction_lift_pushup}), we give a $0.5$ scaling factor to all limbs of the blue character, for the \textit{Salsa Grasping} motion (Figure ~\ref{fig:scaled_heavy_interaction_salsa_grasp}), we scale the yellow character's limbs by $0.5$, and for \textit{Salsa Support} motion (\ref{fig:scaled_heavy_interaction_salsa_support}), we scale the yellow character's limbs by $0.8$. For this type of motion, our method allows the scaled characters to adjust their motions to preserve the interactions rather than simply mimicking the original reference motions, and therefore the semantics of the interaction are transferred successfully to the scaled characters. For example, the taller characters in the \textit{Lift-Pushup} motions learned to bend down and reach the target grasping region to form the grasp.

 
Finally, we also scale the characters and objects for human-object interaction scenarios. Figure \ref{fig:human_object_interaction} shows the control policies learned successfully for the small box throwing-catching and the large box lifting-moving. For both human-object interaction motions, we scale the yellow character's limbs by $0.7$. \jungdam{Need some explanation on how the characters and objects are scaled (scale factors) + Other tricks if exist.}\yunbo{Included all scaling factors. Need to modify the text when referring to characters in the figure after we have the rendered figures.}

\subsection{Non-human Characters}
\begin{figure}[H]
    \centering
    \includegraphics[width=0.3\textwidth]{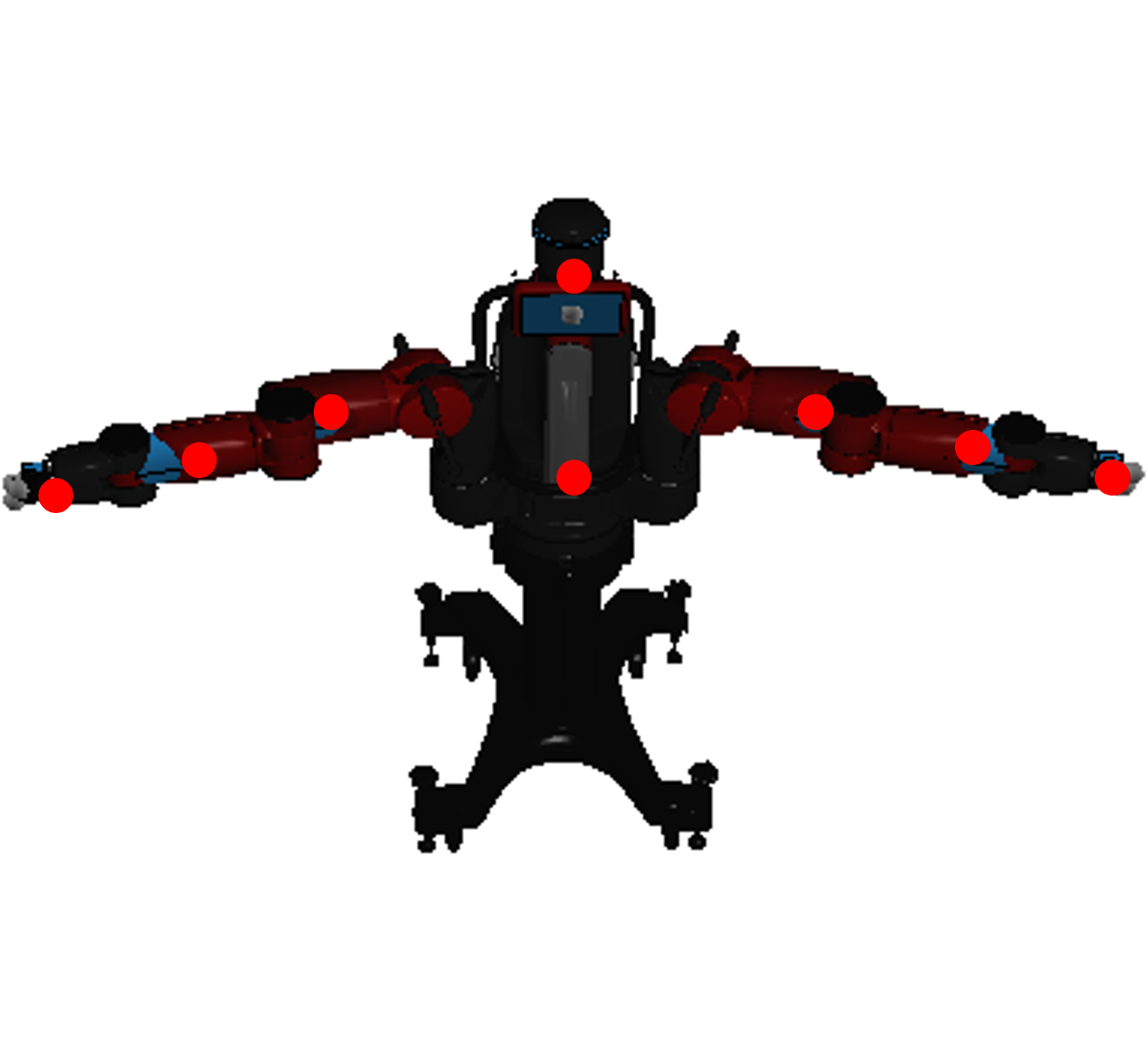}
    \caption{Baxter robot with markers}
    \label{fig:baxter_robot}
\end{figure}

Our method can also transfer interactions in the reference motions to characters with different kinematic configurations. For example, if we  use a robot with fewer DoFs than the reference character, our method can still make the robot create the interactions existing in the reference motions. As shown in Figure \ref{fig:baxter_robot}, we replace one of the characters by a Baxter robot composed of two industrial manipulators. 

Because the robot has a fixed base, we place the robot at the location where the greeting motion is conducted and place a total of eight markers on the upper body of the robot on its head, torso, upper arms, lower arms, and end-effectors to match those of the human character. For the human character, we keep the same 15 markers as described earlier on the human body. We then use a total of 23 markers to construct the interaction graph for the training. During the training, we use two separate reward functions for the character and robot. The character receives the same reward terms as described above, the robot only receives a reward from $r_{pos\_graph}$ and $r_{vel\_graph}$ because it is not mobile.

In addition, we found that including the first term in Equation~\ref{eq:edge_weighting_function} was helpful for the robot because it was immobile. This term highlights the edge error when the robot body parts are staying close but the reference characters' body is far away. \yunbo{Mention we use the bidirectional version of the edge-weighting function}  

\begin{figure}[H]
     \begin{subfigure}{0.45\columnwidth}
        \centering
        \includegraphics[height=3.5cm]{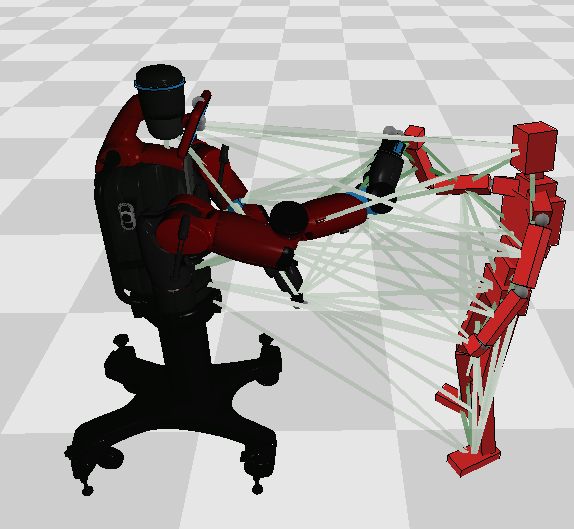}
        \caption{Human-Baxter Greeting}
        \label{fig:human_baxter_greetings}
         \end{subfigure}
     \begin{subfigure}{0.45\columnwidth}
        \centering
        \includegraphics[height=3.5cm]{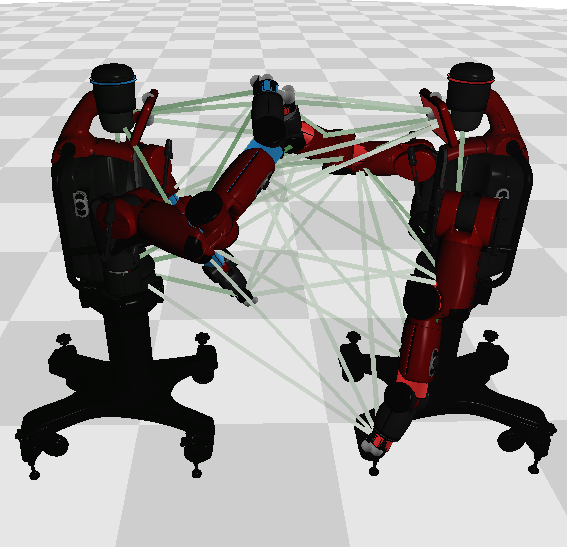}
         \caption{Baxter-Baxter Highfive}
        \label{fig:baxter_baxter_highfive}
     \end{subfigure}
    \caption{Robot Interaction}
    \label{fig:robot_interaction}
\end{figure}

Because the kinematic structure of the robot is completely different that of the actor in the reference character/motion, we ask the policy to directly output the absolute target joint angles $q$ instead of learning the deviation (i.e. $\Delta q$) from the reference $q_{ref}$ for both the human character and the robot. Our framework can successfully generate animations of the Baxter robot performing greetings with a human character (Figure ~\ref{fig:human_baxter_greetings}), and perform a highfive with another Baxter robot (Figure ~\ref{fig:baxter_baxter_highfive}). These examples demonstrate the potential of our method as an option to retarget human motion onto robots and create compelling human-robot interactions.


\subsection{Comparison}
We conduct comparison and ablation studies to show the effectiveness of our graph-based formulation in reproducing complex interaction for physically simulated characters. 

\subsubsection{Joint-based Reward}
\begin{figure}[H]
    \centering
     \begin{subfigure}{0.45\columnwidth}
        \centering
        \includegraphics[height=3.5cm]{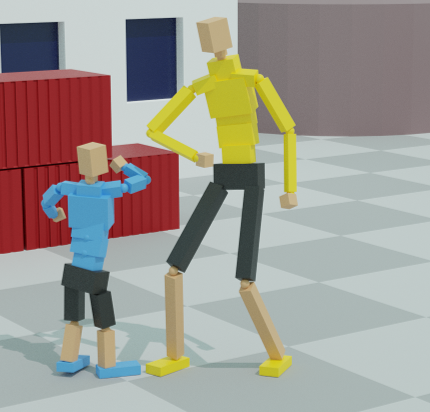}
        \caption{Joint-based Reward}
        \label{fig:scadiver_greetings_comparison_scadiver}
     \end{subfigure}
     \begin{subfigure}{0.45\columnwidth}
        \centering
        \includegraphics[height=3.5cm]{figs/different_scale/IG_scaled_greetings.png}
        \caption{Interaction Graph Reward}
        \label{fig:scadiver_greetings_comparison_IG}
     \end{subfigure}
     \caption{Reward comparison for Greetings.}
    \label{fig:scadiver_greetings_comparison}
\end{figure}

To highlight the necessity of formulating the interaction-graph-based reward, we compare our method with the commonly used joint-based reward formulation for motion imitation. For the sequences that use a joint-based reward, we apply a similar formulation as described in \cite{Peng:2018:DeepMimic, Won:2020:ScaDiver}. That formulation asks the policy to minimize the positional and angular differences of the joints and links between the simulation and the reference motion. In this formulation, no reward term exists to evaluate the quality of interactions between multiple characters or characters to objects. As a result, when the simulated character has a different body configuration from the reference motion, the characters will only learn to mimic the poses in the reference motion instead of learning to adapt to the other characters (or object) to correctly perform the interaction. Figure \ref{fig:scadiver_greetings_comparison} shows a comparison for the greeting motions. The control policies trained using a joint-based reward fail to cause the taller character to bend down to meet the shorter character. Similar behaviors are observed in the other motions for the control policies trained using the joint-based reward only.

\begin{figure}[H]
    \centering
     \begin{subfigure}{0.45\columnwidth}
        \centering
        \includegraphics[height=3.5cm]{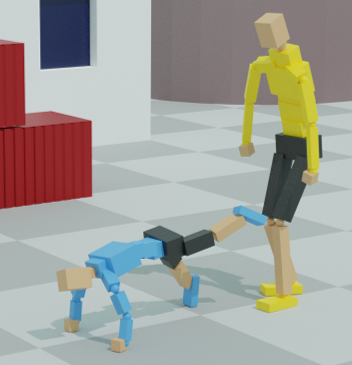}
        \caption{Joint-based Reward}
        \label{fig:scadiver_pushup_comparison_scadiver}
     \end{subfigure}
     \begin{subfigure}{0.45\columnwidth}
        \centering
        \includegraphics[height=3.5cm]{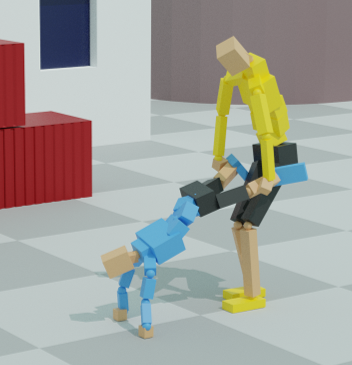}
        \caption{Interaction Graph Reward}
        \label{fig:scadiver_pushup_comparison_IG}
     \end{subfigure}
     \caption{Reward comparison for Lift Pushup.}
      \label{fig:scadiver_pushup_comparison}
\end{figure}

We further contrast the performance for the dense interaction example between the interaction graph  and joint-based rewards. Figure~\ref{fig:scadiver_pushup_comparison} shows such a comparison on \textit{Lift-Pushup} sequence with a scaled character. When using the interaction graph reward, the taller character actively bends forward to reach its hands to the shorter character's lower leg to form the grasping constraints and lift the shorter character. When using a joint-based reward, on the other hand, there is no reward based on the relative poses between the two characters and the taller character cannot grasp the shorter character's leg and the interaction semantics are not preserved. 

\begin{figure}[H]
    \centering
     \begin{subfigure}{0.45\columnwidth}
        \centering
        \includegraphics[height=3.5cm]{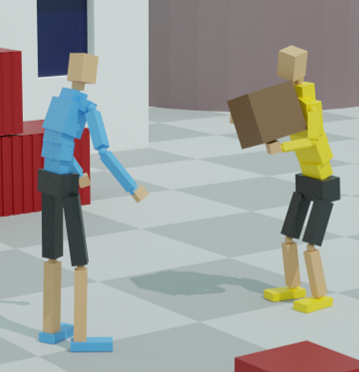}
        \caption{Joint-based Reward}
        \label{fig:scadiver_box_comparison_scadiver}
     \end{subfigure}
     \begin{subfigure}{0.45\columnwidth}
        \centering
        \includegraphics[height=3.5cm]{figs/different_scale/IG_Scaled_Box_Throw.png}
        \caption{Interaction Graph Reward}
        \label{fig:scadiver_box_comparison_IG}
     \end{subfigure}
     \caption{Box Throw comparison}
      \label{fig:scadiver_box_comparison}
\end{figure}

Furthermore, we show that a joint-based reward also produces lower quality motions when re-targeting motions for human-object interactions. Figure~\ref{fig:scadiver_box_comparison} shows a comparison for a small box throw and catch motion trained with the interaction graph reward and joint-based reward. The two characters are able to perform the throw and catch motion sequence in the joint-based reward because of the presence of the additional object observation and reward as described above. However, it fails to preserve the interaction semantics because the shorter character should catch the box by holding on two opposite faces of the box instead of supporting the box on its bottom. 

\subsubsection{Edge Weighting Function}
\begin{figure}[H]
    \centering
     \begin{subfigure}{0.45\columnwidth}
        \centering
        \includegraphics[height=3.5cm]{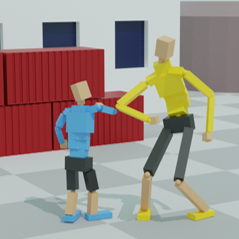}
        \caption{Without Edge Weights}
        \label{fig:weighting_function_ablation_without}
     \end{subfigure}
     \begin{subfigure}{0.45\columnwidth}
        \centering
        \includegraphics[height=3.5cm]{figs/different_scale/IG_scaled_greetings.png}
        \caption{With Edge Weights}
        \label{fig:weighting_function_ablation_with}
     \end{subfigure}

     \caption{Comparison showing the effect of edge weights}
      \label{fig:weighting_function_ablation}
\end{figure}

We do an ablation on the edge weighting function (Equation~\ref{eq:edge_weighting_function}) to understand how this helps the training selectively pay attention to more important edges and ignore irrelevant edges. Our experiments demonstrate that this design can help in generating more natural-looking motions.  In Figure~\ref{fig:weighting_function_ablation}, we compare the resulting policy trained with (left) and without (right) the weighting function for the greeting motion. When the edge weighting function is present, the taller character learns to bend its waist to reduce its height when greeting the shorter character. However, when all the edges have the same weight during training, the taller character instead learns to walk and complete all the greetings with the legs bent at an unnatural angle. This unnatural behavior is created because the policy tries to get a low error on every edge of the graph regardless of the distances of the nodes.
\section{Discussion}

We demonstrated a method of simulating and retargeting complex multi-Character interactions by using deep reinforcement learning where novel state and rewards that are character-agnostic are developed based on an \textit{Interaction Graph}. Our formulation is applicable to a variety of interactions among people ranging from sparse interactions (e.g. greeting, jumpover) to complex ones (e.g. exercise motion, Salsa dancing) regardless of whether the body size, kinematics or skeleton of the simulated characters are the same as that of the actors who recorded the reference motions. 

While we demonstrate many successful examples, there are some limitations to our method. First, there are some limitations of our reward function design. Because the action space of our policy is not directly associated with the reward function, our training usually requires more samples to converge compared to a joint-based reward function. In addition, due to the lack of supervision on the joint angles, the motion generated from our policy could contain artifacts on joints that have little impact to the interaction. For example, sometimes the character may tilt the head or the waist at an unnatural angle because this deviation from the reference will not affect the positions of the interaction graph's node, and therefore it does not decrease the reward. Adding more markers would be an immediate remedy but this would also increase computational cost. Another limitation is that our controllers are imitation controllers which cannot perform interactions that do not exist in the reference motions.  Further the controllers only work for the specific body configuration that it was trained on, so one policy cannot easily be generalized to work on a character with a different body configuration. \limVar{ We also observe that the variability of our result is limited by the dissimilarity of the character and the difficulty of the task. Extreme scaling or drastically different skeletons could fail to imitate the interactions due to their physical limits. For example, in challenging interaction scenarios such as box throwing, our method fails when replacing one human character with a robot.}

We envision several future directions \limVar{to reduce the limitations}. \limVar{For better variability, we can build a stronger motion prior that contains a larger variety of motion types. Further training on top of the motion prior could be more sample efficient, and allow the policy to explore the motion space to find a valid solution when the character shape undergoes extreme changes.} \limGen{ To improve the generalization of our method, a better observation representation would be helpful. Currently we are using the commonly used joint-based observation representation for the policy to account for the environmental context. A better formulation of the observation space that directly interprets the interaction graph (instead of joint angles) would be a first step towards a universal policy. A good candidate would be utilizing a \textit{Graph Neural Network} to directly observe the interaction graph in the simulation to better understand the spatial relationship of different objects in the scene. After the creation of a comprehensive model that can better understand the interaction observations, a subsequent future direction would be building a interaction prior model that can sense the surrounding environment and create motions to complete general interactions without requiring specific reference motions as demonstrated in~\cite{Peng:2022:ASE, Won:2022:cVAE} for a single character motion. We believe that our method has made it possible to generate complex multi-character interactions of simulated characters for many behaviors and will serve as a stepping stone for future research. }


\begin{acks}
Jungdam Won was partially supported by the New Faculty Startup Fund from Seoul National University, ICT(Institute of Computer Technology) at Seoul National University, and the MSIT (Ministry of Science and ICT), Korea, under the ITRC (Information Technology Research Center) support program (IITP-2023-2020-0-01460) supervised by the IITP(Institute for Information \& Communications Technology Planning \& Evaluation).
\end{acks}

\bibliographystyle{ACM-Reference-Format}
\bibliography{bibliography}
\clearpage
\newpage

 






\end{document}